%Paper: hep-th/9503009
%From: oda@ins.u-tokyo.ac.jp
%Date: Thu, 02 Mar 95 18:15:17 +0900

\input phyzzx%
\hsize=417pt %\vsize=600pt \baselineskip=20pt \maxdepth=0.2pt
\sequentialequations
\Pubnum={ EDO-EP-2 }
\date={ February 17, 1995 }
\titlepage
\vskip 32pt
\title{ Thermodynamics of Black Hole in (N+3)-dimensions from Euclidean N-brane
Theory}
\author{Ichiro Oda\footnote \dag{E-mail addess:
sjk13904@mgw.shijokyo.or.jp}}
\vskip 16pt
\address{ Edogawa University,
          474 Komaki, Nagareyama City,
          Chiba 270-01, JAPAN     }

%
%              The titlepage ends at this place.
%
%======================================================================%
%
\abstract{In this article we consider an N-brane description of
an (N+3)-dimensional black hole horizon. First of
all, we start by reviewing a previous work where
a string theory is used as describing the dynamics
of the event horizon of a four dimensional black
hole. Then we consider a particle model defined on one
dimensional Euclidean line in a three dimensional black hole as a
target spacetime metric. By solving the field equations
 we find a ``world line instanton'' which connects
the past event horizon with the future one. This solution gives us the
exact value of the Hawking temperature and to leading order
 the Bekenstein-Hawking formula of black hole entropy.
We also show that this formalism is extensible to an arbitrary
spacetime dimension.
  Finally we make a comment of one-loop quantum
 correction to the black hole entropy .}
\endpage
%
%=========================================================================%
%
%        Macros
%

\def\sp(#1){\noalign{\vskip #1pt}}

%
%        REFERENCES
%
\REF\Hawking{
          S.W.Hawking,
          Comm. Math. Phys. {\bf 43} (1975) 199.
          }
\REF\Beken{
          J.D.Bekenstein,
          Nuovo Cim. Lett. {\bf 4} (1972) 737 ;
          Phys. Rev. {\bf D7} (1973) 2333 ;
          ibid. {\bf D9} (1974) 3292;
          Physics Today {\bf 33}, no.1 (1980)  24.
          }
\REF\OdaI{
          I.Oda,
          Phys. Lett. {\bf B338} (1994) 165.
         }
\REF\tHooft{
          G.'t Hooft,
          Nucl. Phys. {\bf B335} (1990) 138;
          Physica Scripta {\bf T15} (1987) 143;
          ibid. {\bf T36} (1991) 247.
          }
\REF\OdaII{
          I.Oda,
          Int. J. Mod. Phys. {\bf D1} (1992) 355.
          }
\REF\Unruh{
          W.G.Unruh,
          Phys. Rev.{\bf D14} (1976) 870.
          }
\REF\Birrell{
          N.D.Birrell and P.C.Davies, Quantum fields in curved space (Cambridge
U.P.,
          Cambridge, 1982).
          }
\REF\Gibb{
          G.Gibbons and S.W.Hawking, Phys. Rev. {\bf D49}
          (1977) 2752; S.W.Hawking, From General Relativity:
          An Einstein Centenary Survey (Cambridge U.P.,
          Cambridge, 1979).
          }
\REF\Banados{
          M.Ba$\tilde n$ados, C.Teitelboim and J.Zanneli, Phy. Rev.
          Lett. {\bf 69} (1992) 1849.
          }
\REF\Dabho{
          A.Dabholkar, Harvard preprint HUTP-94-A019, hep-th/9408098;
          Caltech preprint CALT-68-1953, hep-th/9409158.
          }
\REF\Lowe{
          D.A.Lowe and A.Strominger, Santa Barbara preprint UCSBTH-94-42,
          hep-th/9410215.
          }
\REF\Sergei{
          E.Elizalde and S.D.Odintsov, Barcelona preprint UB-ECM-PF 94/30,
          hep-th/9411024.
          }
\REF\Emp{
          R.Emparan, preprint EHU-FT-94/10, hep-th/9412003.
          }
%
%
%
%
%=========================================================================%
%
%   This part is the meat of body.
%
\topskip 30pt
\par
The Bekenstein-Hawking formula of the black hole entropy,
$S={1 \over 4} {k c^3 \over G \hbar} A_H${ \PRrefmark{\Hawking,\Beken}
is not only so beautiful but also very mysterious for us at present.
 This formula contains
the four fundamental constants of physics, those are, the Boltzman constant
$k$, the Newton one $G$,
the Planck one $\hbar$ and the light velocity $c$ so that it suggests
a deep triangle relation among thermodynamics, general relativity and
quantum mechanics. Moreover, this formula relates the entropy of a black
hole to the area of a event horizon, therefore also implies a connection to
geometry. Thus, although the above formula was originally derived
in terms of the semi-classical approach, many theoretical physicists never
doubt its validity up to some quantum corrections
even when we have a quantum theory of gravitation in the future.

However, the underlying physical basis by which ${1 \over 4} {k c^3 \over G
\hbar} A_H$ arises as the black hole entropy remains unclear.
It is tempted to regard this black entropy as the logarithm of the number
of microscopic states compatible with the observed macroscopic state from
the viewpoint of the ordinary statistical physics. Then, a crux of
an understanding is what those microscopic states are.

It might be true that the underlying law explaining the
Bekenstein-Hawking formula might presumably not be fully
understood until we construct a theory of quantum gravity. But there
certainly exists an opposite attitude toward it. Namely this
mystery might give us a clue of constructing a theory of quantum
gravity. At this point one expects that a quantum black hole plays
a similar role as the hydrogen atom at the advent of the quantum mechanics.

In a previous paper \PRrefmark{\OdaI}, we have considered
a stringy description of a black hole horizon in four spacetime
dimensions where we have described
the event horizon of a black hole in terms of the world sheet swept
by a string in the Rindler background. It was shown that a nonlinear
sigma model action leads to both the Hawking temperature of a black hole
and the well-known Bekenstein-Hawking formula of the black hole entropy
within the lowest order of approximation. Furthermore we have derived
a covariant operator algebra on the event horizon which is a natural
generalization to the 'tHooft one \PRrefmark{\tHooft,\OdaII}.

Thus it is natural to ask whether this stringy approach to the black hole
thermodynamics can be extended to an arbitrary spacetime dimension or is
a peculiar thing only in four dimensions.
We will see later that we can in fact construct a more general formalism
where the event horizon of a black hole in (N+3)-spacetime dimensions
is described by a Euclidean N-brane.

We start with a brief review of the previous work on a string, i.e.,
1-brane, approach to the four dimensional black hole \PRrefmark{\OdaI}.
As an effective action describing the dynamical properties of
the black hole horizon, let us consider the Polyakov action of
a bosonic string in a curved target spacetime which is given by
\par
%%%%%%%%%%%%%%%%%%%%%%%%%%%%Equation%%%%%%%%%%%%%%%%%%%%%%%%%%%%%%%%%%%
$$ \eqalign{ \sp(2.0)
S_{(1)} &= -{T \over 2} \int d^2 \sigma \sqrt{h}
        h^{\alpha\beta} \partial_{\alpha}X^{\mu} \partial_{\beta}
        X^{\nu} g_{\mu\nu}(X),
\cr
\sp(3.0)} \eqno(1)$$
%-----------------------------------------------------------------------
where $T$ is a string tension having dimensions of mass squared,
$h_{\alpha\beta}(\tau,\sigma)$ denotes the two dimensional world
sheet metric having a Euclidean signature, and $h =det h_{\alpha
\beta}$. $X^{\mu}(\tau,\sigma)$ maps the string into four dimensional
spacetime, and then $g_{\mu\nu}(X)$ can be identified as the background
spacetime metric in which the string is propagating. Note that $\alpha,
\beta$ take values 0, 1 and $\mu,\nu$ do values 0, 1, 2, 3.

The classical field equations from the action (1) become
%%%%%%%%%%%%%%%%%%%%%%%%%%%%Equation%%%%%%%%%%%%%%%%%%%%%%%%%%%%%%%%%%%
$$ \eqalign{ \sp(2.0)
0 &= T_{\alpha\beta} = -{2 \over T} {1 \over \sqrt{h}}
     {\delta S_{(1)} \over \delta h^{\alpha\beta}},
\cr
  &= \partial_{\alpha}X^{\mu} \partial_{\beta}X^{\nu}
     g_{\mu\nu}(X) - {1 \over 2} h_{\alpha\beta} h^{\rho\sigma}
     \partial_{\rho}X^{\mu} \partial_{\sigma}X^{\nu}
     g_{\mu\nu}(X),
\cr
\sp(3.0)} \eqno(2)$$
%-----------------------------------------------------------------------
%%%%%%%%%%%%%%%%%%%%%%%%%%%%Equation%%%%%%%%%%%%%%%%%%%%%%%%%%%%%%%%%%%
$$ \eqalign{ \sp(2.0)
0 &= \partial_{\alpha} (\sqrt{h} h^{\alpha\beta} g_{\mu\nu}
     \partial_{\beta}X^{\nu}) - {1 \over 2} \sqrt{h}
     h^{\alpha\beta} \partial_{\alpha}X^{\rho} \partial_{\beta}
     X^{\sigma} \partial_{\mu} g_{\rho\sigma}.
\cr
\sp(3.0)} \eqno(3)$$
%-----------------------------------------------------------------------
Here let us consider the case where the background spacetime metric
$g_{\mu\nu}(X)$ takes a form of the Euclidean Rindler metric
%%%%%%%%%%%%%%%%%%%%%%%%%%%%Equation%%%%%%%%%%%%%%%%%%%%%%%%%%%%%%%%%%%
$$ \eqalign{ \sp(2.0)
ds^2 = g_{\mu\nu}dX^{\mu}dX^{\nu}
     =+g^2z^2dt^2 + dx^2 + dy^2 + dz^2,
\cr
\sp(3.0)} \eqno(4)$$
%-----------------------------------------------------------------------
where $g$ is given by $g = {1 \over 4M}$. It is well-known that
the Rindler metric can be obtained in the large mass limit
from the Schwarzschild black hole
metric and provides us with a nice playground for examining
the Hawking radiation in a simple metric form
\PRrefmark{\Unruh, \Birrell}. Here it is important to notice that we
have performed the Wick rotation with respect to the time component
since now we would like to discuss the thermodynamic properties
of the Rindler spacetime when we assume that the dynamics of the
event horizon is controlled by a Euclidean string.

Now one can easily solve Eq.(2) as follows:
%%%%%%%%%%%%%%%%%%%%%%%%%%%%Equation%%%%%%%%%%%%%%%%%%%%%%%%%%%%%%%%%%%
$$ \eqalign{ \sp(2.0)
h_{\alpha\beta} = G (\tau,\sigma) \partial_{\alpha}
                  X^{\mu} \partial_{\beta}X^{\nu}
                  g_{\mu \nu}(X),
\cr
\sp(3.0)} \eqno(5)$$
%-----------------------------------------------------------------------
where $G(\tau,\sigma)$ denotes the Liouville mode. Next we shall fix
the gauge symmetries which are the two dimensional diffeomorphisms
and the Weyl rescaling by
%%%%%%%%%%%%%%%%%%%%%%%%%%%%Equation%%%%%%%%%%%%%%%%%%%%%%%%%%%%%%%%%%%
$$ \eqalign{ \sp(2.0)
x(\tau,\sigma) = \tau, \ y(\tau,\sigma) = \sigma, \
                 G(\tau,\sigma) = 1.
\cr
\sp(3.0)} \eqno(6)$$
%-----------------------------------------------------------------------
At this stage, let us impose a cyclic symmetry
%%%%%%%%%%%%%%%%%%%%%%%%%%%%Equation%%%%%%%%%%%%%%%%%%%%%%%%%%%%%%%%%%%
$$ \eqalign{ \sp(2.0)
z(\tau,\sigma) = z(\tau), \ t(\tau,\sigma) = t(\tau).
\cr
\sp(3.0)} \eqno(7)$$
%-----------------------------------------------------------------------
Some people might be afraid that this additional assumption kills
various physically interesting states by which one cannot derive
the entropy formula, but we will see that such a situation never
occurs at least at the lowest order of approximation. Incidentally,
we will not impose the same kind of symmetry ansatz in the case of
three dimensional black hole. From Eqs.(5), (6) and (7), the world sheet
metric takes the form
%%%%%%%%%%%%%%%%%%%%%%%%%%%%Equation%%%%%%%%%%%%%%%%%%%%%%%%%%%%%%%%%%%
$$ \eqalign{ \sp(2.0)
h_{\alpha\beta} = \left(\matrix{g^2 z^2 \dot t^2
                  + \dot z^2 + 1 & 0 \cr
                  0 & 1 \cr} \right),
\cr
\sp(3.0)} \eqno(8)$$
%-----------------------------------------------------------------------
where the dot denotes a derivative with respect to $\tau$.
And the remaining field equations (3) become
%%%%%%%%%%%%%%%%%%%%%%%%%%%%Equation%%%%%%%%%%%%%%%%%%%%%%%%%%%%%%%%%%%
$$ \eqalign{ \sp(2.0)
\partial_{\tau}
({{z^2 \dot t} \over \sqrt{h}}) = 0,
\cr
\sp(3.0)} \eqno(9)$$
%-----------------------------------------------------------------------
%%%%%%%%%%%%%%%%%%%%%%%%%%%%Equation%%%%%%%%%%%%%%%%%%%%%%%%%%%%%%%%%%%
$$ \eqalign{ \sp(2.0)
\partial_{\tau}h = 0,
\cr
\sp(3.0)} \eqno(10)$$
%-----------------------------------------------------------------------
%%%%%%%%%%%%%%%%%%%%%%%%%%%%Equation%%%%%%%%%%%%%%%%%%%%%%%%%%%%%%%%%%%
$$ \eqalign{ \sp(2.0)
\partial_{\tau}
( \ {\dot z \over \sqrt{h}} \ )
- {1 \over \sqrt{h}}g^2 z {\dot t}^2 = 0,
\cr
\sp(3.0)} \eqno(11)$$
%-----------------------------------------------------------------------
where
%%%%%%%%%%%%%%%%%%%%%%%%%%%%Equation%%%%%%%%%%%%%%%%%%%%%%%%%%%%%%%%%%%
$$ \eqalign{ \sp(2.0)
h = g^2 z^2 \dot t^2 + \dot z^2 + 1.
\cr
\sp(3.0)} \eqno(12)$$
%-----------------------------------------------------------------------

Now it is straightforward to solve the above field equations.
We have two kinds of solutions. One is a trivial one given by
%%%%%%%%%%%%%%%%%%%%%%%%%%%%Equation%%%%%%%%%%%%%%%%%%%%%%%%%%%%%%%%%%%
$$ \eqalign{ \sp(2.0)
z = \dot z = \ddot z = 0, t(\tau) = arbitrary.
\cr
\sp(3.0)} \eqno(13)$$
%-----------------------------------------------------------------------
The other is the solution of ``world sheet instanton'' described by
%%%%%%%%%%%%%%%%%%%%%%%%%%%%Equation%%%%%%%%%%%%%%%%%%%%%%%%%%%%%%%%%%%
$$ \eqalign{ \sp(2.0)
z(\tau) = \sqrt{c_2 (\tau - \tau_0)^2 +
          {g^2 c_1^2 \over c_2}},
\cr
t - t_0 = {1 \over g} {\tan}^{-1}{c_2 \over {gc_1}}
           (\tau - \tau_0),
\cr
\sp(3.0)} \eqno(14)$$
%-----------------------------------------------------------------------
where $c_1, c_2, \tau_0$, and $t_0$ are the integration
parameters, in other words, ``moduli parameters''.
In order to understand the physical meaning of this solution
more vividly, it is convenient to eliminate the variable
$\tau$ and express $z$ in terms of the time variable $t$.
Then from Eq.(14) we obtain
%%%%%%%%%%%%%%%%%%%%%%%%%%%%Equation%%%%%%%%%%%%%%%%%%%%%%%%%%%%%%%%%%%
$$ \eqalign{ \sp(2.0)
z(t_E) \ = \ {g c_1 \over \sqrt{c_2}}
           {1 \over \cos{g(t_E - t_{E0})}},
\cr
\sp(3.0)} \eqno(15)$$
%-----------------------------------------------------------------------
where we inserted the suffix $E$ on $t$ in order to indicate the
Euclidean time clearly. Furthermore after Wick-rerotating, we have
in the real Lorentzian time $t_L$
%%%%%%%%%%%%%%%%%%%%%%%%%%%%Equation%%%%%%%%%%%%%%%%%%%%%%%%%%%%%%%%%%%
$$ \eqalign{ \sp(2.0)
z(t_L) \ = \ {g c_1 \over \sqrt{c_2}}
           {1 \over \cosh{g(t_L - t_{L0})}}.
\cr
\sp(3.0)} \eqno(16)$$
%-----------------------------------------------------------------------
Here note that the Rindler coordinate $(z,t)$ is related to the
Minkowski coodinate $(Z,T)$ by the following transformation:
%%%%%%%%%%%%%%%%%%%%%%%%%%%%Equation%%%%%%%%%%%%%%%%%%%%%%%%%%%%%%%%%%%
$$ \eqalign{ \sp(2.0)
Z = z \cosh gt, \ T = t \sinh gt,
\cr
\sp(3.0)} \eqno(17)$$
%-----------------------------------------------------------------------
thus the above solution (16) corresponds to
an instanton connecting the past event horizon with the future one with a
definite constant value $Z = {g c_1 \over \sqrt{c_2}}$.

Now let us examine the thermodynamic properties of the ``world sheet
instanton''. It is remarkable that the solution has a periodicity
with respect to the Euclidean time component, $\beta = {2 \pi \over g}$
whose inverse gives us nothing but the Hawking temperature $T_H =
{1 \over \beta} = {g \over 2\pi} = {1 \over 8 \pi M}$ of the Rindler
spacetime \PRrefmark{\Unruh}. Next let us calculate the black
hole entropy to the leading order of approximaion by a method developed
by Gibbons and Hawking \PRrefmark{\Gibb}. The result is
%%%%%%%%%%%%%%%%%%%%%%%%%%%%Equation%%%%%%%%%%%%%%%%%%%%%%%%%%%%%%%%%%%
$$ \eqalign{ \sp(2.0)
S = \sqrt{c_2 + 1} \ T \ A_H,
\cr
\sp(3.0)} \eqno(18)$$
%-----------------------------------------------------------------------
where $A_H = \int dx dy$ which corresponds to the area of the black
hole horizon. At this stage, by selecting the string tension
%%%%%%%%%%%%%%%%%%%%%%%%%%%%Equation%%%%%%%%%%%%%%%%%%%%%%%%%%%%%%%%%%%
$$ \eqalign{ \sp(2.0)
T = {1 \over 4 \sqrt{c_2 + 1}G},
\cr
\sp(3.0)} \eqno(19)$$
%-----------------------------------------------------------------------
we arrive at the famous Bekenstein-Hawking entropy formula \PRrefmark
{\Hawking, \Beken}
%%%%%%%%%%%%%%%%%%%%%%%%%%%%Equation%%%%%%%%%%%%%%%%%%%%%%%%%%%%%%%%%%%
$$ \eqalign{ \sp(2.0)
S = {1 \over 4G} A_H.
\cr
\sp(3.0)} \eqno(20)$$
%-----------------------------------------------------------------------

Let us apply the formalism mentioned so far for a three dimensional
black hole model \PRrefmark{\Banados} whose line element is given by
%%%%%%%%%%%%%%%%%%%%%%%%%%%%Equation%%%%%%%%%%%%%%%%%%%%%%%%%%%%%%%%%%%
$$ \eqalign{ \sp(2.0)
ds^2 = -(- M + {r^2 \over l^2}) dt^2
       + (- M + {r^2 \over l^2})^{-1} dr^2 + r^2 d\phi^2,
\cr
\sp(3.0)} \eqno(21)$$
%-----------------------------------------------------------------------
where $l^2$ and the cosmological constant $\Lambda$ have a relation
like $\Lambda = -{1 \over l^2}$. From Eq.(21), one can have the Rindler
metric in the limit of a small cosmological constant
after an appropriate change of variables. In fact, defining
%%%%%%%%%%%%%%%%%%%%%%%%%%%%Equation%%%%%%%%%%%%%%%%%%%%%%%%%%%%%%%%%%%
$$ \eqalign{ \sp(2.0)
x = l \sqrt{M} \phi, \ z = {l \over \sqrt{M}}
    \sqrt{- M + {r^2 \over l^2}},
\cr
\sp(3.0)} \eqno(22)$$
%-----------------------------------------------------------------------
and identifying $g = {\sqrt{M} \over l}$, one has the following
Rindler metric
%%%%%%%%%%%%%%%%%%%%%%%%%%%%Equation%%%%%%%%%%%%%%%%%%%%%%%%%%%%%%%%%%%
$$ \eqalign{ \sp(2.0)
ds^2  = -g^2z^2dt^2 + dx^2  + dz^2.
\cr
\sp(3.0)} \eqno(23)$$
%-----------------------------------------------------------------------
Note that the black hole metric Eq.(21)
and the Rindler metric Eq.(23) differ only by terms $O(({z \over l})^2).$

In case of the three dimensional Rindler spacetime, we need to
consider a particle (0-brane) model defined on one dimensional
Euclidean line with the Euclidean Rindler metric as a spacetime metric.
Thus we start by an action
%%%%%%%%%%%%%%%%%%%%%%%%%%%%Equation%%%%%%%%%%%%%%%%%%%%%%%%%%%%%%%%%%%
$$ \eqalign{ \sp(2.0)
S_{(0)} &= -{T \over 2} \int d \tau
          ( \ {1 \over e} \dot X^{\mu} \dot X^{\nu} g_{\mu\nu}(X)
          + e \ ),
\cr
\sp(3.0)} \eqno(24)$$
%-----------------------------------------------------------------------
where $e$ is the einbein. Note that now $T$ must have dimensionality
of mass to leave a dimensionless action. Of course, in this case,
$\mu,\nu$ take values 0, 1, 2.

{}From Eq.(24), one has the field equations through the variation
of the einbein and $X^{\mu}$
%%%%%%%%%%%%%%%%%%%%%%%%%%%%Equation%%%%%%%%%%%%%%%%%%%%%%%%%%%%%%%%%%%
$$ \eqalign{ \sp(2.0)
0 &= \dot X^{\mu} \dot X^{\nu} g_{\mu\nu}(X) - e^2,
\cr
\sp(3.0)} \eqno(25)$$
%-----------------------------------------------------------------------
%%%%%%%%%%%%%%%%%%%%%%%%%%%%Equation%%%%%%%%%%%%%%%%%%%%%%%%%%%%%%%%%%%
$$ \eqalign{ \sp(2.0)
0 &= \partial_{\tau} ( \ {1 \over e} \ g_{\mu\nu}
     \dot X^{\nu} \ ) - {1 \over 2 e}
     \dot X^{\rho} \dot X^{\sigma} \partial_{\mu} g_{\rho\sigma}.
\cr
\sp(3.0)} \eqno(26)$$
%-----------------------------------------------------------------------
Now we shall fix the gauge symmetry which is one dimensional
diffeomorphism by
%%%%%%%%%%%%%%%%%%%%%%%%%%%%Equation%%%%%%%%%%%%%%%%%%%%%%%%%%%%%%%%%%%
$$ \eqalign{ \sp(2.0)
x(\tau) = \tau.
\cr
\sp(3.0)} \eqno(27)$$
%-----------------------------------------------------------------------
{}From Eqs.(25), (26) and (27), it is easy to show that
the field equations can be rewritten
into a perfectly equivalent form to Eqs.(9)-(12) by defining
$h = e^2$.

In a completely analogous way to the string case, we can obtain
the ``world line instanton'' solution Eq.(15). This solution
gives us the Hawking temperature $T_H = {1 \over \beta} =
{g \over 2\pi} = {\sqrt{M} \over 2 \pi l}$ of the three
dimensional black hole \PRrefmark{\Banados}. Now the entropy
is given by
%%%%%%%%%%%%%%%%%%%%%%%%%%%%Equation%%%%%%%%%%%%%%%%%%%%%%%%%%%%%%%%%%%
$$ \eqalign{ \sp(2.0)
S = \sqrt{c_2 + 1} \ T \ L_H,
\cr
\sp(3.0)} \eqno(28)$$
%-----------------------------------------------------------------------
where $L_H = \int dx$ denotes the length of the black hole horizon.
Since both the particle coupling constant $T$ and the inverse of
the Newton constant $G$ in three dimensions have a dimension
of mass, one can set up the relation (19) and then reach the
Bekenstein-Hawking entropy formula (20) where $A_H$ is now
replaced with $L_H$.

Let us generalize the present formalism to the black hole
in (N+3)-dimensions where we need to consider the Euclidean
N-brane. We will assume $N \geq 2$ without generality.
Our starting action as an effective action describing
the dynamics of the event horizon is given by
%%%%%%%%%%%%%%%%%%%%%%%%%%%%Equation%%%%%%%%%%%%%%%%%%%%%%%%%%%%%%%%%%%
$$ \eqalign{ \sp(2.0)
S_{(N)} &= -{T \over 2} \int d^{N+1} \sigma \sqrt{h}
        ( \ h^{\alpha\beta} \partial_{\alpha}X^{\mu} \partial_{\beta}
        X^{\nu} g_{\mu\nu}(X) + 1 - N \ ),
\cr
\sp(3.0)} \eqno(29)$$
%-----------------------------------------------------------------------
where $T$ is an N-brane tension having dimensionality
$(mass)^{N+1}$. Now $\alpha,
\beta$ take values 0, 1,..., N and $\mu,\nu$ do values 0, 1,..., N+2.
The variation of the action (29) gives us the field equations
%%%%%%%%%%%%%%%%%%%%%%%%%%%%Equation%%%%%%%%%%%%%%%%%%%%%%%%%%%%%%%%%%%
$$ \eqalign{ \sp(2.0)
0 &= T_{\alpha\beta} = -{2 \over T} {1 \over \sqrt{h}}
     {\delta S_{(N)} \over \delta h^{\alpha\beta}},
\cr
  &= \partial_{\alpha}X^{\mu} \partial_{\beta}X^{\nu}
     g_{\mu\nu}(X) - {1 \over 2} h_{\alpha\beta}
     ( \ h^{\rho\sigma}
     \partial_{\rho}X^{\mu} \partial_{\sigma}X^{\nu}
     g_{\mu\nu}(X) + 1 - N \ ),
\cr
\sp(3.0)} \eqno(30)$$
%-----------------------------------------------------------------------
and Eq.(3). The Schwarzschild black hole metric in (N+3)-spacetime
dimensions takes a form in a Euclidean signature
%%%%%%%%%%%%%%%%%%%%%%%%%%%%Equation%%%%%%%%%%%%%%%%%%%%%%%%%%%%%%%%%%%
$$ \eqalign{ \sp(2.0)
ds^2 = +( 1 - {2M \over r}) dt^2
       + ( 1 - {2M \over r})^{-1} dr^2 + r^2 d\Omega^2_{N+1},
\cr
\sp(3.0)} \eqno(31)$$
%-----------------------------------------------------------------------
which in the large mass limit can be written to the Rindler metric
%%%%%%%%%%%%%%%%%%%%%%%%%%%%Equation%%%%%%%%%%%%%%%%%%%%%%%%%%%%%%%%%%%
$$ \eqalign{ \sp(2.0)
ds^2  =+g^2z^2dt^2 + dz^2 + \sum_{i=1}^{N+1} dx^2_i,
\cr
\sp(3.0)} \eqno(32)$$
%-----------------------------------------------------------------------
with $g = {1 \over 4M}$.

As before, one can solve Eq.(30) as follows:
%%%%%%%%%%%%%%%%%%%%%%%%%%%%Equation%%%%%%%%%%%%%%%%%%%%%%%%%%%%%%%%%%%
$$ \eqalign{ \sp(2.0)
h_{\alpha\beta} =  \partial_{\alpha}
                  X^{\mu} \partial_{\beta}X^{\nu}
                  g_{\mu \nu}(X).
\cr
\sp(3.0)} \eqno(33)$$
%-----------------------------------------------------------------------
The gauge symmetries, which are (N+1)-dimensional diffeomorphisms,
are fixed by the gauge conditions
%%%%%%%%%%%%%%%%%%%%%%%%%%%%Equation%%%%%%%%%%%%%%%%%%%%%%%%%%%%%%%%%%%
$$ \eqalign{
x_1(\tau, \sigma_1, \cdots, \sigma_N)&=\tau, \cr
x_2(\tau, \sigma_1, \cdots, \sigma_N)&=\sigma_1, \cr
\ldots \ldots \cr
x_{N+1}(\tau, \sigma_1, \cdots, \sigma_N)&=\sigma_N. \cr
\sp(3.0)} \eqno(34)$$
%-----------------------------------------------------------------------
Furthermore we shall make a cyclic ansatz
%%%%%%%%%%%%%%%%%%%%%%%%%%%%Equation%%%%%%%%%%%%%%%%%%%%%%%%%%%%%%%%%%%
$$ \eqalign{
z(\tau, \sigma_1, \cdots, \sigma_N)&=z(\tau), \cr
t(\tau, \sigma_1, \cdots, \sigma_N)&=t(\tau). \cr
\sp(3.0)} \eqno(35)$$
%-----------------------------------------------------------------------
{}From Eqs.(33), (34) and (35), the world volume metric becomes
%%%%%%%%%%%%%%%%%%%%%%%%%%%%Equation%%%%%%%%%%%%%%%%%%%%%%%%%%%%%%%%%%%
$$ \eqalign{
    h_{\alpha\beta}=\pmatrix{
   g^2z^2 \dot t^2 + \dot z^2 + 1&0&\ldots&0 \cr
   0&1&\ldots&0 \cr
   \vdots&\vdots &\ddots&\vdots \cr
   0&0&\ldots&1 \cr}.
} \eqno(36)$$
%-----------------------------------------------------------------------
\par

As the remaining field equations we have the same equations as Eqs.(9)-(12)
,thus as an interesting classical solution ``world volume instanton''
Eq.(15). Of course this solution has a periodicity whose inverse is
exactly equal to the Hawking temperature, and the entropy is given
in terms of the Bekenstein-Hawking formula Eq.(20) where this time
$A_H$ denotes the volume of the event horizon given by
$\int \prod_{i=1}^{N+1}dx_i$.

Finally let us make a comment on one-loop quantum correction to
the black hole entropy. Recently there have appeared several
articles where string theories have been utilized
 \PRrefmark{\Dabho,\Lowe,\Sergei,\Emp}. An idea behind these
researches is that string theories might provide a finite
quantum correction to the Bekenstein-Hawking entropy formula
owing to a mild ultraviolet behavior at a short distance scale.
However, from the viewpoint of the present work, just in
four spacetime dimensions one has to use Euclidean string theories
in order to understand the quantum mechanical meaning of
the black hole entropy. We would like to consider this problem
in the near future.

\vskip 1cm

\noindent

{\bf Acknowledgments}

	The author would like to thank G.'tHooft for valuable discussions
	and a kind hospitality at Utrecht University.
\refout
\vfill
%=======================================================================%
%
\bye